# Approaching Zero-Temperature Metallic States

# in Mesoscopic Superconductor-Normal-Superconductor Arrays

Serena Eley[1], Sarang Gopalakrishnan[1], Paul M. Goldbart[2], Nadya Mason[1]

**Systems of superconducting islands placed on normal metal films offer tunable realizations of two-dimensional (2D) superconductivity[1,2]; they can thus elucidate open questions regarding the nature of 2D superconductors and competing states. In particular, island systems have been predicted to exhibit zero-temperature metallic states[3-5]. Although evidence exists for such metallic states in some 2D systems[6,7], their character is not well understood: the conventional theory of metals cannot explain them[8], and their properties are difficult to tune[7,9]. Here, we characterize the superconducting transitions in mesoscopic island-array systems as a function of island thickness and spacing. We observe two transitions in the progression to superconductivity; both transition temperatures exhibit unexpectedly strong depression for widely spaced islands. These depressions are consistent with the system approaching zero-temperature metallic states. The nature of the transitions and the state between them is explained using a phenomenological model involving the stabilization of superconductivity on each island via a weak coupling to and feedback from its neighbors.**

Conventional zero-temperature ($T=0$) metallic states do not exist in 2D systems possessing any disorder, because of Anderson localization[8,9]. To reconcile this fact with experimental evidence for $T=0$ metals in 2D, it has been proposed that the experimental observations do not pertain to conventional metals, but rather to spatially inhomogeneous superconducting (or, more generally, correlated) states[3,4,10]. Inhomogeneity is thought to arise in some of these systems because of phase

[1] Department of Physics and Frederick Seitz Materials Research Laboratory, University of Illinois Urbana-Champaign, Urbana, IL 61801, USA
[2] School of Physics, Georgia Institute of Technology, 837 State Street, Atlanta, GA 30332, USA

separation; however, it can also be tunably engineered, e.g., in hybrid superconductor-normal-superconductor (SNS) systems, such as the arrays studied here. In arrays of SNS junctions, the diffusion of electron pairs from the superconductor into the normal metal[11-13]—known as the proximity effect—gives rise to global superconductivity, through a transition typically described using the phenomenological theory of Lobb, Abraham, and Tinkham (LAT)[14]. According to the LAT theory, the $T=0$ state is always superconducting.

Most previous studies of SNS arrays utilized islands much larger than the superconducting coherence length $\xi_{SC}$ (i.e., having well-defined superconductivity)[1]; however, there is evidence that arrays of *mesoscopic* islands (i.e., islands of dimensions comparable to $\xi_{SC}$) exhibit behavior that deviates from the LAT theory[5,15]. In addition, the dependence of the superconducting transition on key parameters—such as island spacing and size—has not been studied systematically. In this Letter, we present transport measurements on arrays of mesoscopic niobium (Nb) islands having systematically varying inter-island spacings, placed on patterned gold (Au) films. We observe that the device resistance drops to zero in two steps as the temperature is lowered. The lower-temperature drop, at temperature $T_2$, is associated with superconducting phase-locking across the array; the dependence of $T_2$ on island spacing and thickness deviates strongly from LAT theory. Surprisingly, the higher-temperature drop, $T_1$, traditionally associated with the superconducting transition of each island, also depends strongly on the island spacing. The pronounced decrease in $T_1$ for larger island spacings indicates that the superconductivity on each individual island is in fact fragile, and that a $T=0$ metallic state might be realizable for very weakly-coupled islands.

Our samples consist of 10 nm-thick Au, patterned for four-point transport measurements, on Si/SiO$_2$ substrates (see Methods for fabrication details). The Au patterns are overlaid with triangular arrays of 260 nm diameter Nb islands, as shown in Fig. 1. Each array contains more than 10,000 Nb islands. The data in this Letter are from two sets of devices: the first has 87 nm (± 2 nm) thick Nb islands, and the second has 145 nm (± 2 nm) thick ones. The devices in each set are identical, except for varied island spacing. X-ray diffraction and scanning electron microscopy of the Nb revealed columnar grains ~ 30 nm in diameter, typical of evaporated Nb[16]; thus, each island contains ~ 50-100 grains. The superconducting coherence length of Nb is estimated to be ~ 27 nm (see Methods), comparable to the grain size but smaller than the island size.

Figure 2 shows resistance measurements for the devices, as well as an illustration of the two-step development of superconductivity. The data in Figs. 2a and 2b show that both $T_1$ and $T_2$ decrease with increasing island spacing. Note that $T_1$ decreases most rapidly for the shorter islands, but does not saturate to a fixed value for either sample as the spacing increases. In addition, the resistance exhibits an abrupt change in slope at $T_1$, but not the sharp drop seen for larger islands[1]. In contrast to previous results[2], $T_2$ is also more strongly depressed for shorter islands. As schematized in Fig. 2c, these phenomena can be understood using a model of coupled islands, each composed of grains, having two characteristic energy scales: (i) $J$, the coupling between grains on an individual island, and (ii) $J'$ $(< J)$, the coupling between grains on neighboring islands. According to this scheme, for $T > T_1$, the separate grains on each island have incoherent superconducting phases; at $T_1$, *intra*-island phase coherence develops, and the system's resistance decreases. For very large islands, $T_1$ would depend only on $J$, which grows with island height but is spacing-independent. For mesoscopic islands, however, the $T_1$ of an *isolated* island is depressed (possibly to $T = 0$) by phase fluctuations among the grains; the inter-island coupling $J'$ serves to reduce these fluctuations by increasing the effective "dimensionality" of the island system, thereby stabilizing superconductivity. Thus, $T_1$ decreases for larger spacings (i.e., as $J'$ decreases). Below $T_1$, the *intra*-island phase coherence strengthens continuously (Fig. 2c, Region II); thus, the system resistance continuously decreases rather than steeply dropping at $T_1$. Region III of Fig. 2c shows the familiar proximity behavior; here, the normal-metal coherence length[17] $\xi_N$ increases until it becomes comparable to the island spacing. Then, *inter*-island phase coherence begins to emerge (Fig. 2c, Region IV), and at $T_2$ the system undergoes a Berezinskii-Kosterlitz-Thouless transition to a fully superconducting state[1, 2].

The inset to Fig. 3 shows how $T_2$ decreases with island spacing. For each device, $T_2$ was extracted by measuring the temperature at which *I-V* curves became non-linear (see Supplementary Information). The dependence of $T_2$ on array parameters deviates from LAT theory both *quantitatively* (i.e., $T_2$ decreases more rapidly with island spacing than predicted) and *qualitatively* (i.e., $T_2$ depends strongly on island *height*). Figure 3 also shows the systematic dependence of $\xi_N(T_2)$ on island spacing, where $\xi_N = \sqrt{\hbar D/(k_B T)}$ and the normal-metal diffusion constant $D \approx 94$ cm$^2$/s (see Methods). We observe $\xi_N(T_2)$ to vary approximately linearly with island spacing.

We now turn to a more quantitative description of these transitions. From the Ginzburg-Landau perspective, $T_1$ for an isolated island of lateral dimensions exceeding $\xi_{SC}$ should equal the transition temperature for a continuous film of the same height, because the suppression of superconductivity due to superconductor-vacuum interfaces at the sides is negligible[13,18]. This expectation is inconsistent with the data. Because the islands are themselves granular, we attribute the unexpected depression of $T_1$ to spatial fluctuations of the superconducting phase *within* each island, which we model using the following Hamiltonian:

$$H = -J \sum_p \sum_{\langle ij \rangle \in p} \cos(\theta_i - \theta_j) - J' \sum_{\langle pp' \rangle} \left( \sum_{i \in p} \cos(\theta_i) \right) \left( \sum_{j \in p'} \cos(\theta_j) \right) \qquad (1)$$

where $\theta_i$ is the superconducting phase of grain $i$, $p$ indexes islands, $\langle ij \rangle \in p$ denotes nearest-neighbor grains on island $p$, and $\langle pp' \rangle$ denotes nearest-neighbor islands. Each grain on an island is assumed to couple with equal strength to every grain on neighboring islands; hence the inter-island interaction $J'$ can be regarded as "mean-field." The temperature dependence of $J'$ is taken to have the standard proximity form[2, 14]: $J'(T) \approx J_0' \exp(-d/\xi_N(T))$, where $d$ is the edge-to-edge spacing of the islands and $J_0'$ is the coupling amplitude. To explain the experimental results the model must have two properties: (i) strong, non-saturating depression of $T_1$ for widely spaced islands, and (ii) a value of $T_1$ that is nevertheless greater than the inter-island coherence temperature $T_2$. We consider a simple model possessing these features, where each island is treated as a chain of XY spins. By exploiting a mapping between the XY chain and a quantum mechanical rotor[19], one can show that the threshold for an island to acquire a well-defined superconducting phase in the mean field of neighboring islands is:

$$\frac{k_B T_1}{\sqrt{zJJ'}} \coth\left(m\sqrt{\frac{zJ'}{J}}\right) = 1. \quad (2)$$

where $z$ is the coordination number of each island (six for a triangular array) and $m$ is the number of grains on each island. This model satisfies property (i) because Eq. (2) allows for $T_1 = 0$ at $J' = 0$ (i.e., isolated islands are not superconducting). The model also satisfies property (ii) as follows. For temperatures well below $T_1$, the phases of the grains on an individual island are mutually locked; hence, one can neglect the first term of Eq. (1). Then, $T_2$ is given by $zm^2 J'(T_2) \approx k_B T_2$. Comparing expressions for $T_1$ and $T_2$, we find two separate transitions (i.e, $T_2 < T_1$) provided that $m < \sqrt{J/J'}$; this condition always holds for large $J$. Although one-dimensional coupled spins are by no means a fully realistic model of the experimental system, this model does capture the main features of the data, and therefore likely the basic physics of the system. Moreover, if we take $J$ to have the proximity-effect form $J \approx J_0 \exp(-\alpha/\xi_N)$, with $\alpha$ being a constant that is weakly dependent on individual-island parameters, then Eq. (2) can be rearranged to yield the following dependence of $\xi_N$ on $d$:

$$d + \alpha = -\ln\left[J_0 J_0' / \xi_N^2(T_1)\right]\xi_N(T_1). \quad (3)$$

Figure 4 shows that fits of the data to Eq. (3) are reasonable. Note that the lack of saturation of $T_1$ seen in the data and reproduced by the model is consistent with a value of $T_1 = 0$; this admits the possibility of a $T = 0$ metallic state, similar to that described in Ref. [4].

We now turn to the spacing- and height-dependence of $T_2$. As seen in Fig. 3, $\xi_N(T_2)$ depends approximately linearly on $d$; such a relationship implies, in particular, that for large $d$ the transition occurs when $d/\xi_N(T_2)$ is a constant. This observation conflicts with the LAT theory[2, 14] (presumed to be valid for our measurement regime of $d > \xi_N$), which predicts that $k_B T_2 \sim J_0' \exp[-d/\xi_N(T_2)]$ (see fits in Fig. 3). The asymptotic constancy of $d/\xi_N(T_2)$ can be accounted for in one of two ways. The first is to modify the LAT theory by replacing the proximity expression for $J'$ with the quasiclassical $T=0$ expression[12] $J' \sim 1/d^2$. This replacement, strictly valid only for $d \leq \xi_N(T_2)$, would yield a modified LAT threshold of the form $T_2 \sim 1/d^2$, and consequently a linear relationship between $d$ and $\xi_N(T_2)$. Although this modification of LAT theory would explain the observed linear relationship, it does not explain the height-dependence of $T_2$. An alternative explanation, which captures both the linear relationship and height dependence, is to assume the existence of an effective "charging energy" $\kappa$ for each island, while retaining the proximity-effect form of $J'$. In this scenario, $T_2$ would occur when $J'_0 \exp[-d/\xi_N(T_2)] = \kappa$; this implies the existence of a minimum inter-island coupling $\kappa$ that must be overcome for superconductivity to be attained, even at $T=0$, and consequently the possibility of a $T=0$ metallic state.

Both phase transitions studied in this work occur at temperatures that seem to extrapolate to zero, suggesting the existence of two quantum phase transitions; in particular, our devices may approach a quantum superconductor-metal transition, which has been predicted but not observed[3-5, 10]. The tunability of our devices—including the ability to vary island geometry, spacing, material properties, and disorder—thus makes them excellent test-beds for exploring such transitions. The unconventional metallic state we observe at temperatures between $T_1$ and $T_2$ is similar to that predicted for phase-separated quantum metals[4, 5] in that it possesses "regional" phase correlations, i.e., correlations at length-scales larger than that of single islands but not global in extent. The ability to stabilize regional correlations, in the absence of long-range ordering, is characteristic of a variety of inhomogeneous correlated systems, including high-temperature superconductors[20], coupled magnetic chains[21], and strained superconducting films[22]. The tunability of our system could thus help elucidate open questions in these materials.

**Methods**

**Samples.** Standard photolithographic techniques and electron beam evaporation were used to create the 10-nm thick four-point pattern of Au with a 4 Å Ti adhesion layer. The Nb islands were then patterned using electron-beam lithography. Prior to electron beam evaporation of the Nb islands( in a UHV system at ~$10^{-9}$ torr), the Au surface was $Ar^+$ ion milled to establish a clean interface. For each sample, six arrays were patterned onto a single $Si/SiO_2$ substrate, each with different edge to edge spacings: $d$ = 90 nm, 140 nm, 190 nm, 240 nm, 290 nm, and 340 nm. Four other samples were fabricated; all showed similar data trends, but had limited data ranges (e.g. fewer working devices).

The measurement area of every device is 120 $\mu m$ x 30 $\mu m$, so the number of islands in each array ranges from 11,400 to 33,516, depending on the island spacing. The large number of islands ensures that discrete percolation paths or individual junction properties do not dominate the conductance. All Nb islands are 260 nm in diameter, which is about 10 times the Nb Ginzburg-Landau dirty-limit coherence length $\xi_{SC}^{Nb}(T_c \approx 9.1K) \approx 27$nm, for an approximate mean free path $l \approx 8$nm. We estimate $l$ from the Einstein relation, using the normal state resistivity $\rho \approx 1.12 \times 10^{-5} \Omega \cdot cm$ near the transition of an 87 nm thick unpatterned Nb film. X-ray diffraction spectra of Nb films and scanning electron microscopy of Nb islands showed that the Nb is polycrystalline with growth along the (110) direction, and grain height equivalent to the film thickness. SEM images revealed an elongated, columnar grain structure. The Au film resistivity in all devices is $\rho(10K) \approx (6.25 \pm 0.75) \times 10^{-6} \Omega \cdot cm.$ Using the Einstein relation, we estimate a diffusion constant $D \approx 94.2 cm^2/s,$ which yields a mean free path of $l \approx$ 13 nm and a temperature-dependent coherence length $\xi_N(T) \approx 268/\sqrt{T}$ nm.

**Measurement.** All measurements above 1.5 K were carried out in a pumped He-4 cryostat, while lower temperature measurements were performed in a He-3 refrigerator. Resistance was measured by standard, low-frequency ac lock-in techniques using an excitation current of 500 nA. To minimize Joule heating, *IV* characteristics were measured using rectangular current pulses, with a current-on time of 3.5 ms and current-off time of 3 ms.

Acknowledgements

This research was supported by the DOE-DMS under grant DE-FG02-07ER46453 through the Frederick Seitz Materials Research Laboratory at the University of Illinois at Urbana-Champaign, and partly carried out in the MRL Central Facilities (partially supported by the DOE under DE-FG02-07ER46453 and DE-FG02-07ER46471).

Author Contributions

S.E. carried out the experimental work. S.G. and P.M.G. carried out the theoretical analysis. All authors discussed the results and commented on the manuscript.

**Figures**

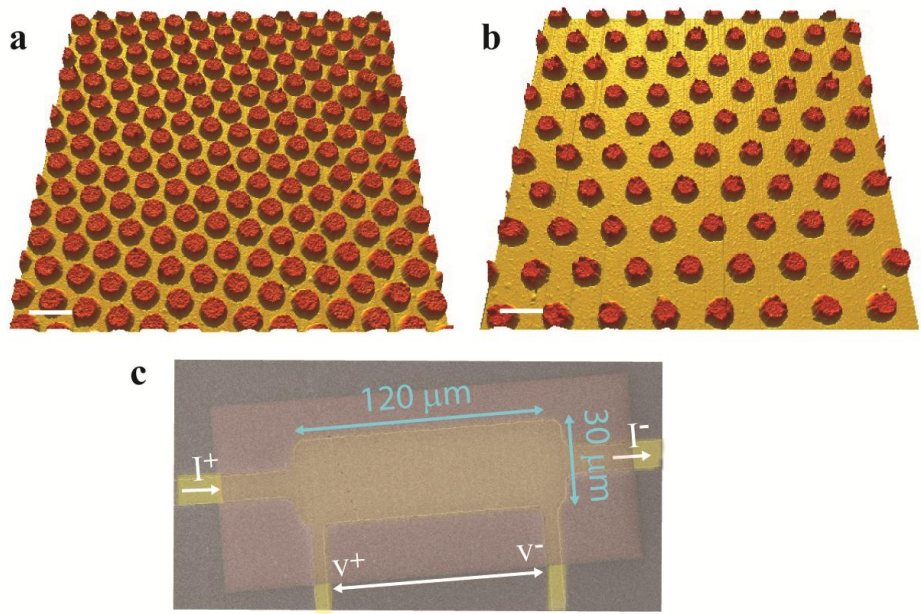

**Figure 1| AFM topography of arrays of Nb islands on Au and SEM image of device. a, b** AFM images of arrays of 87-nm thick Nb islands (red) on 10-nm thick Au underlayer (yellow). Each array has an edge-to-edge spacing of **a,** 140 nm and **b,** 340 nm. The white scale bar is 500 nm. **c,** False color SEM image of island array (red rectangle) overlapping Au four-probe pattern (yellow), with the measurement schematic indicated.

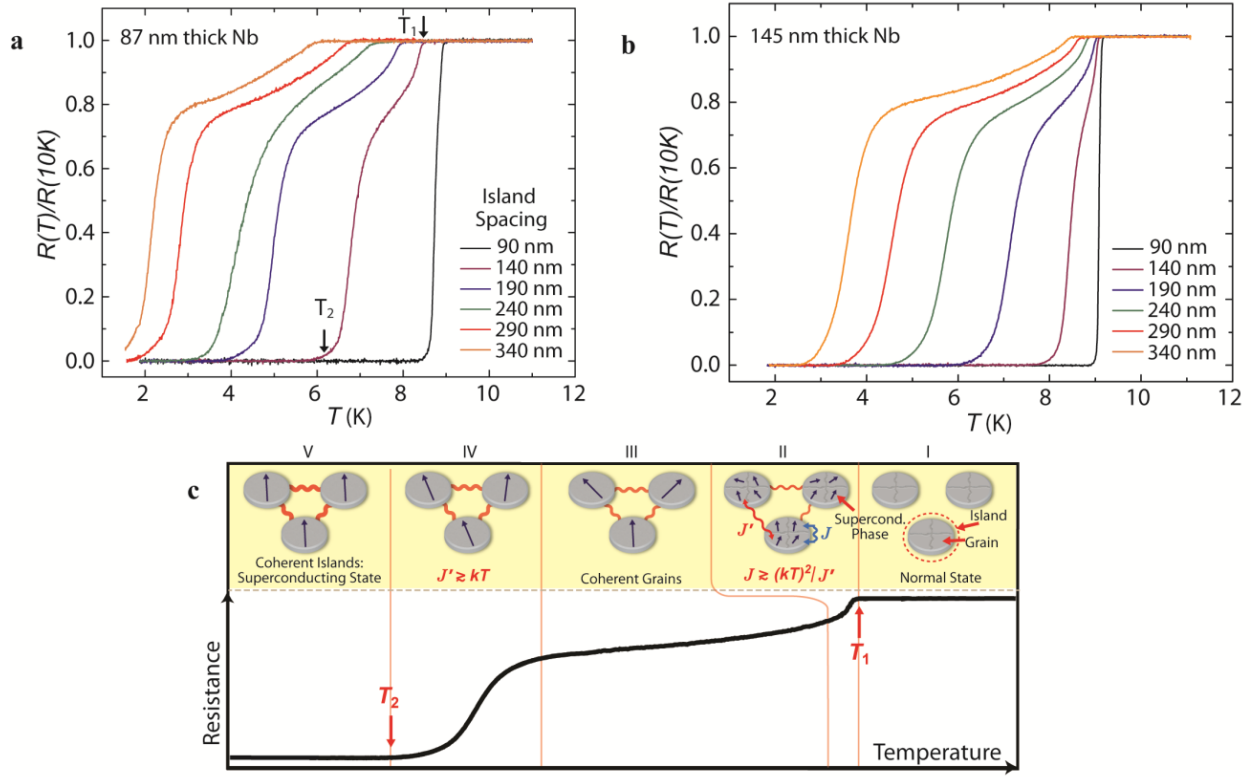

**Figure 2| Superconductivity in Nb island arrays. a, b** Temperature dependent resistive transitions in arrays having different edge-to-edge island spacings. The island diameter is 260 nm for all arrays. The islands are **a,** 87 nm thick and **b**, 145 nm thick. Black arrows in **a** mark $T_1$ and $T_2$ for the islands spaced 140-nm apart. The data are normalized to the resistance at 10K. Note that $T_1$ and $T_2$ occur at higher temperatures for thicker islands. In Panel **a**, the lowest temperature curves are cut off by the minimum attainable temperature of our apparatus. **c,** Curve illustrates two-step $R$ vs. $T$ behavior, with the island transition marked at $T_1$, and film transition marked at $T_2$. Pictures show three islands, each limited to four grains for simplicity. In region I, the Nb islands are normal metals. In region II, the phase of the *grains* (represented by arrows) starts to become coherent throughout each island (although there is not yet inter-island phase coherence). At $T_1$, Cooper pairs diffuse from the Nb into the Au, and the resistance drops. The grains have *intra*-island Josephson coupling $J$ and nearest-neighbor *inter*-island coupling $J'$(represented by red squiggly lines). In region III, $J$ has saturated, but $J'$ continues to increase as the normal metal coherence length $\xi_N$ increases. In region IV, $\xi_N$ becomes comparable to the island spacing, and the entire system of film and islands progresses toward having global phase coherence. As the temperature is further decreased, the film undergoes a transition to a superconducting state at $T_2$.

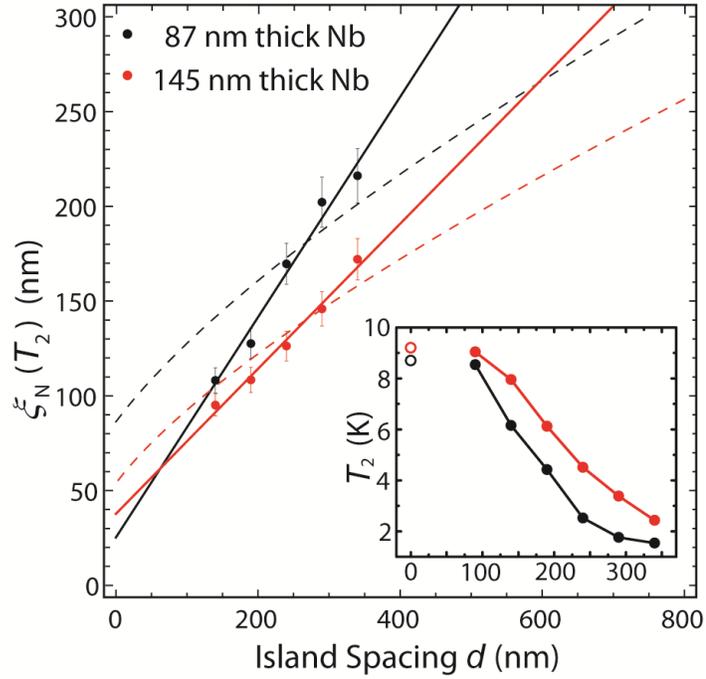

**Figure 3| Dependence of $T_2$ on array geometry. Main Panel,** The normal-metal coherence length at $T_2$, i.e. $\xi_N(T_2)$, is shown as a function of island spacing. The temperature $T_2$ for each device was extracted from the temperature-dependence of *IV* curves (see Supplementary Information). Red and black lines are fits to linear curves, and dotted lines are fits to LAT theory (Ref. 14). The point for the closest spaced islands is excluded from the plot, as for them the transition shows only one step. The **inset** shows $T_2$ for each device vs. edge-to-edge spacing, for 87 nm-thick islands (black dots) and 145 nm-thick islands (red dots). The open circles mark the $T_c$ of the un-patterned bilayers (8.75 K and 9.1 K, respectively).

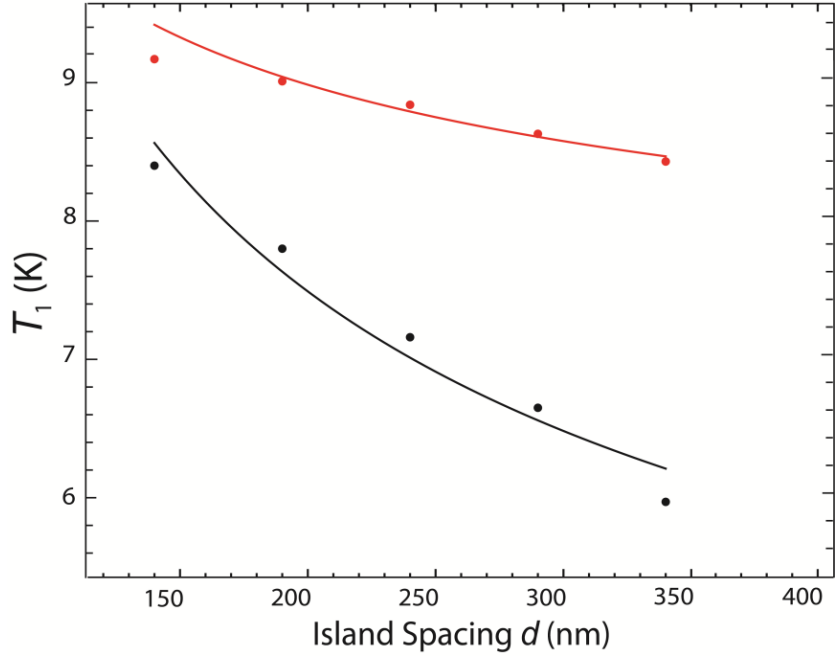

**Figure 4|** Dependence of the first transition temperature $T_1$ on the island spacing for 87-nm thick islands (black points) and 145-nm thick islands (red points). Solid lines are fits to Eq. (3), i.e., the coupled-XY-chain model, in which the coupling $J_0$ and the length scale $\alpha$ are treated as fit parameters (see Text). $J_0$ depends strongly on island height, but $\alpha$ is approximately constant. The points for the smallest island spacings are not shown, as the transitions for this spacing do not show two steps; for a similar reason, the 140-nm spacing for the thicker islands is shown, but not included in the fit. A notable discrepancy involves the largest (340-nm) spacing for the thinner islands, for which $T_1$ is lower than the model predicts; this might suggest that other—possibly quantum—fluctuations are significant in this regime.

# Supplementary Information: Approaching Zero-Temperature Metallic States in Mesoscopic Superconductor-Normal-Superconductor Arrays

Serena Eley[3], Sarang Gopalakrishnan[1], Paul M. Goldbart[4], Nadya Mason[1]

[1] *Department of Physics and Frederick Seitz Materials Research Laboratory, University of Illinois Urbana-Champaign, Urbana, IL 61801, USA*
[2] *School of Physics, Georgia Institute of Technology, 837 State Street, Atlanta, GA 30332, USA*

**Extracting $T_2$ from *IV* curves**

The Berezinskii-Kosterlitz-Thouless (BKT) transition temperatures $T_2$ were extracted from *IV* measurements of the devices. SNS arrays are known to undergo a BKT (i.e., vortex-antivortex binding) transition to a fully superconducting state[1]. Below $T_2$, applying a finite excitation current creates a Lorentz force on bound vortex pairs, causing some pairs to unbind. Dissipation caused by current-induced free vortices may overwhelm that caused by thermally unbound vortices near $T_2$, causing a significant resistance at temperatures less than $T_2$. For $T > T_2$, free vortices cause the array to exhibit Ohmic resistance. For $T < T_2$, the *IV* characteristics are nonlinear such that

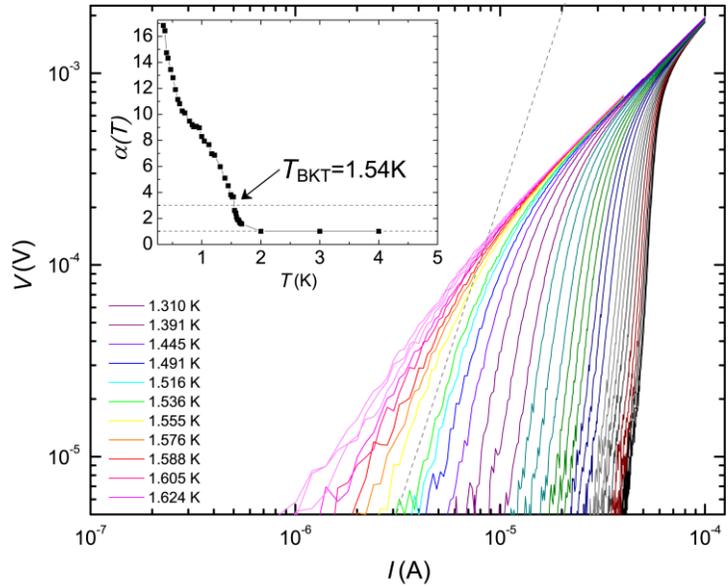

**Identification of BKT Transition. a,** Non-linear *IV* characteristics at different temperatures for 340-nm spaced islands in array with 87 nm Nb, where $V \propto I^{\alpha(T)}$  A slope of $\alpha(T_2) = 3$ is marked by the dotted line. The inset shows the evolution of the slope of the *IV* curves, $\alpha(T)$, and identifies the Nelson-Kosterlitz jump in $\alpha(T)$ at $T_2$.

$V \propto I^{\alpha(T)}$, where $\alpha(T) = 2(T_2/T) + 1$ for $T \leq T_2$ and $\alpha(T) = 1$ for $T > T_2$, leading to $\alpha(T_2) = 3$. This Nelson-Kosterlitz jump in the temperature-dependent exponent $\alpha(T)$ is a universal signature of a BKT transition, although finite-size effects and weak magnetic fields can smear this transition. Consequently, it is standard practice to extract $T_2$ from *IV* characteristics[2]. (In contrast, it is often difficult to accurately fit resistance vs. temperature data to theory, which predicts the flux flow resistance form $R \propto \exp(-b\sqrt{T-T_2})$). For our devices, *IV* curves were measured for a wide range of temperatures above and below $T_2$. All measurements above 1.5 K were performed in a pumped He-4 cryostat, and lower temperature measurements were performed in a He-3 refrigerator. To minimize Joule heating, *IV* characteristics were measured using rectangular current pulses, with a current-on time of 3.5 ms and current-off time of 3 ms. Figure 4 shows a log-log plot of *IV* curves at different temperatures for the 340-nm spaced array with 87-nm thick Nb islands, as an example. The dotted line shows where the slope of the curves is $\alpha(T_2) = 3$, while the inset shows the temperature dependence of α, extracted from the slope of each curve. Note the jump in α(T) at $T_2 = 1.54K$. Results for the other devices showed similar jumps in α(T), which allowed us to extract $T_2$ as a function of array spacing.

---

[1] Resnick, D.J., Garland, J.C., Boyd, J.T., Shoemaker, S. & Newrock, R.S. Kosterlitz-Thouless Transition in Proximity-Coupled Superconducting Arrays. *Physical Review Letters* **47**, 1542 (1981); Abraham, D.W., Lobb, C.J., Tinkham, M. & Klapwijk, T.M. Resistive transition in two-dimensional arrays of superconducting weak links. *Physical Review B* **26**, 5268 (1982).

[2] Nelson, D.R. & Kosterlitz, J.M. Universal Jump in the Superfluid Density of Two-Dimensional Superfluids. *Physical Review Letters* **39**, 1201 (1977).